\documentstyle[nato,epsfig]{crckapb} 

\begin{opening}

\title{On reductions of some KdV-type systems 
and their link to the quartic He'non-Heiles Hamiltonian}    

\subtitle{
Bilinear integrable systems - from classical to quantum, continuous to
discrete,\\
Kluwer.
NATO ARW, 15--19 September 2002, Isola d'Elba. S2003/097.}

\author{C.~Verhoeven$^1$, M.~Musette$^1$, and R.~Conte$^2$}
\institute{
1 Dienst Theoretische Natuurkunde, Vrije Universiteit Brussel\\
Pleinlaan 2, B--1050 Brussels, Belgium\\
E-mail: CVerhoev@vub.ac.be and MMusette@vub.ac.be\\[5pt]
2 Service de physique de l'\'etat condens\'e (URA 2464)\\
CEA--Saclay, F--91191 Gif-sur-Yvette Cedex, France\\
E-mail:  Conte@drecam.saclay.cea.fr
\\ \\ 26~June~2003
}

\end{opening}

\runningtitle{KdV-type systems and He'non-Heiles}          

\begin{document}


\newcommand{\pard}[2]{\frac{\partial #1}{\partial #2}}
\def \ccomma{\raise 2pt\hbox{,}} 
\def \D {\hbox{d}}
\def \Log {\mathop{\rm Log}\nolimits}
\def\PTP{Prog.~Theor.~Phys.~}
\def\CRAS{C.~R.~Acad.~Sc.~Paris}
\def\biSK{bi-SK}
\def\biKK{bi-KK}
\def\biSH{bi-SH}
\def\cKdV{c-KdV}


\begin{abstract}
A few 2+1-dimensional equations belonging to the KP and modi\-fied KP 
hierarchies are shown to be
sufficient to provide a unified picture of all the integrable cases of the
cubic and quartic H\'enon-Heiles Hamiltonians.
\end{abstract}

\section{Introduction}

The H\'enon-Heiles (HH) Hamiltonian \cite{HH} with a generalized cubic
potential is defined as
\begin{eqnarray}
\hspace{-15pt}
\label{HH3:gen}
&&\hbox{HH3}:H =
\frac{1}{2}(p_1^2+p_2^2+c_1q_1^2+c_2q_2^2)+\alpha q_1q_2^2-\frac{\beta}{3}q_1^3
 +\frac{c_3}{q_2^2}\ccomma
\end{eqnarray}
in which $\alpha,\beta,c_1,c_2,c_3$ are constants.

The corresponding equations of motion pass the Painlev\'e test
for only three sets of values of the ratio $\beta / \alpha$,
which are also the only three cases for which an additional first integral $K$
has been found
\cite{BSV:1982,CTW,GDP1982b}.
These three cases have been integrated \cite{Drach1919KdV,VMC2002}
with genus-two hyperelliptic functions.
Moreover, they are equivalent \cite{Fordy1991} to the stationary reduction of
three fifth order soliton equations,
called fifth order Korteweg de Vries (KdV$_5$),
Sawada-Kotera (SK) and Kaup-Kupershmidt (KK) equations,
belonging respectively to the
KP, BKP and CKP hierarchies whose Hirota bilinear forms can be
found in \cite{JM1983}.

If the potential is taken as the most general cubic polynomial in $(q_1,q_2)$,
there exists a fourth Liouville integrable case,
\begin{equation}
V= q_1^3 +\frac{1}{2} q_2^2 q_1 +\frac{i}{6\sqrt{3}} q_2^3\ccomma
\end{equation}
detected by Ramani \textit{et al.} \cite{RDG1982},
but up to now its general solution is unknown.

Another H\'enon-Heiles-type Hamiltonian with an extended quartic potential
has been considered,
\begin{eqnarray}
&&\hbox{HH4}:
H=\frac{1}{2}(P_1^2+P_2^2+a Q_1^2+b Q_2^2)+C Q_1^4+ B Q_1^2 Q_2^2 + A Q_2^4
\nonumber
\\
\label{HH4:gen}
&&\hspace{+80pt}
 +\frac{1}{2}\left(\frac{\alpha}{Q_1^2} +
\frac{\beta}{Q_2^2}\right) + \mu Q_1,
\end{eqnarray}
in which $A,B,C,\alpha,\beta,\mu,a,b$ are constants.
Again,
the equations of motion pass the Painlev\'e test for only
four values of the ratios $A:B:C$ \cite{RDG1982,GDR1983,H1987},
which happen to be the only known cases of Liouville integrability.
However,
it is not yet completely settled whether,
in all four cases,
the quartic Hamiltonian (\ref{HH4:gen}) displays the same pattern as the
cubic Hamiltonian (\ref{HH3:gen}), i.~e.~
\begin{itemize}
\item
the equations of motion can be integrated with
hyperelliptic functions of genus two,
\item
there exists an equivalence with the stationary
reduction of some partial differential equation (PDE)
belonging to the KP, BKP and CKP hierarchies.
\end{itemize}

In this paper,
we first summarize the results already established for
the systems (\ref{HH3:gen}) and (\ref{HH4:gen}).
We then establish new links between the coupled KdV (\cKdV) systems
considered in
\cite{BEF1995b} and some other ones
\cite{DS1984,JM1983,VM2003} belonging to the BKP and CKP hierarchies.
These links could be useful to find the explicit general solution
without any restriction on the parameters other than those generated by the
Painlev\'e test.

\section{Already integrated cases}

The four cases for which the quartic Hamiltonian passes the Painlev\'e test
are,
\begin{description}
\item
(i) $A:B:C= 1:2:1,\mu=0$.
The system is then equivalent to the stationary reduction of the Manakov system
\cite{Manakov1973} of two coupled nonlinear Schr\"odinger (NLS) equations
and has been integrated \cite{Woj1985} with genus two hyperelliptic functions.

\item
(ii)$ A:B:C = 1:6:1, a=b,\mu=0$,

\item
(iii) $A:B:C = 1:6:8, a=4b,\alpha=0$,

\item
(iv) $A:B:C = 1:12:16,a=4b,\mu=0$.

\end{description}
Each of the last three cases is equivalent \cite{BEF1995b}
to the stationary reduction of
a coupled KdV system possessing a fourth or fifth order Lax pair.
Canonical transformations have been found \cite{BEF1995b,BakerThesis}
which allow us \cite{VMC2003,V2003}
to define the separating variables
of the Hamilton-Jacobi equation,
however with additional restrictions on $\alpha,\beta,\mu$,
as showed in Table \ref{Table1}.

{

\begin{table}[h] 
\caption[garbage]{
All the cases of HH3 and HH4 which pass the Painlev\'e test,      
with the extra terms $c_3$ or $\alpha,\beta,\mu$.
First column indicates the cubic or quartic case.
Second column is the value of $\beta/ \alpha$ (if cubic)
or the ratio $A:B:C$ (quartic),
followed by the values selected by the Painlev\'e test.
Third column indicates the polynomial degree of the additional
constant of the motion $K$ in the momenta $(p_1,p_2)$.
Next column displays the PDE system connected to the HH case.
Last column shows the reference to the general solution and
the not yet integrated cases.
When the general solution is known,
it is a singlevalued rational function
of genus two hyperelliptic functions.
}
\vspace{0.2truecm}
\begin{center}
\begin{tabular}{| c | l | c | l | c |}
\hline 
HH
&
case
&
deg $K$
&
PDE
&
General solution
\\ \hline   \hline 
$3$ & $-1,c_1=c_2$         & $4$ & SK      & \cite{VMC2002} 
\\ \hline 
$3$ & $-6$, $c_1,c_2$ arb. & $2$ & KdV$_5$ & \cite{Drach1919KdV}
\\ \hline 
$3$ & $-16$, $c_1=16c_2$   & $4$ & KK      & \cite{VMC2002}
\\ \hline   \hline 
$4$ & $\displaystyle{{1:2:1} \atop {\mu=0}}$ & $2$ & c-NLS & \cite{Woj1985}
\\ \hline 
$4$ & $\displaystyle{{1:6:1} \atop {a=b,\mu=0}}$ & $4$ 
& \cKdV 2, Lax order 4 
& $\alpha=\beta  \cite{VMC2003},\ \alpha\neq\beta ?$
\\ \hline 
$4$ & $\displaystyle{{1:6:8}\atop{a=4b,\alpha=0}}$ & $4$ 
& \cKdV 1, Lax order 4 
& $\beta\mu=0    \cite{VMC2003},\ \beta\mu\neq 0 ?$
\\ \hline 
$4$ & $\displaystyle{{1:12:16} \atop {a=4b,\mu=0}}$ & $4$ 
& \cKdV b Lax order 5 
& $\alpha\beta=0 \cite{V2003},\   \alpha\beta\neq 0 ?$
\\ \hline 
\end{tabular}
\end{center}
\label{Table1}
\end{table}
}

\section{Link between KP hierarchies and integrable HH cases}

\begin{figure}[h]
\centering\epsfig{file=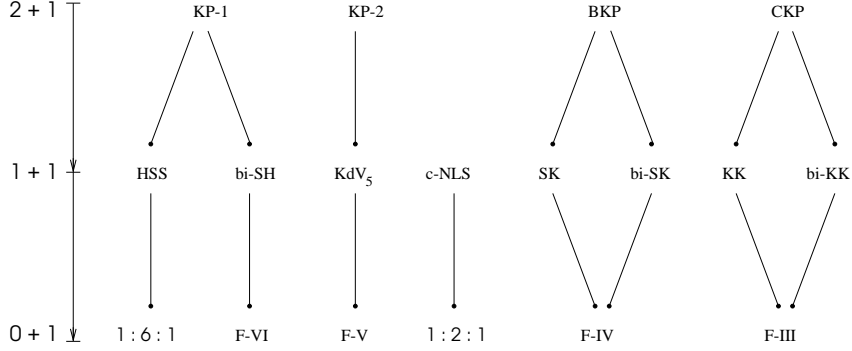,width=0.9\textwidth}
\caption{Reductions from (2+1)-dimensional PDEs to (1+1)-dimensional PDEs,
then to ODEs (the notation F-xxx denotes the autonomous case
of the ODE denoted F-xxx in [6]) 
or to Hamiltonian systems.
The symbol c-NLS represents the Manakov system [18] 
of two coupled NLS equations.}
\label{Fign+1}
\end{figure}

Let us consider the following three systems
of the KP and modified KP hierarchies \cite{JM1983},
\begin{eqnarray}
& &
\left\lbrace
\begin{array}{ll}
\displaystyle{
\left(D_1^4 - 4 D_1 D_3 + 3 D_2^2 \right) (\tau_0\cdot \tau_0) = 0,
}
\\
\displaystyle{
\left(\left(D_1^3 + 2 D_3\right) D_2 - 3 D_1 D_4\right)(\tau_0\cdot \tau_0)=0,
}
\end{array}
\right.
\label{eq:CKPbilinear}
\\
& &
\left\lbrace
\begin{array}{ll}
\displaystyle{
\left(D_1^4 - 4 D_1 D_3  + 3 D_2^2 \right) (\tau_0\cdot \tau_0) = 0,
}
\\
\displaystyle{
\left(D_1^6 - 20 D_1^3 D_3 - 80 D_3^2 + 144 D_1D_5 - 45 D_1^2 D_2^2\right)
 (\tau_0\cdot\tau_0) = 0,
}
\end{array}
\right.
\label{eq:KPbilinear}
\\
& &
\left\lbrace
\begin{array}{ll}
\displaystyle{
\left( D_1^2  + D_2\right)(\tau_0\cdot\tau_1) = 0,
}
\\
\displaystyle{
\left(D_1^6 - 20 D_1^3 D_3 - 80 D_3^2 + 144 D_1 D_5
\right.
}
\\
\displaystyle{
\left.
\phantom{xxxxxxx}
 + 15\left( 4 D_1 D_3 - D_1^4\right) D_2\right)(\tau_0\cdot\tau_1) = 0,
}
\end{array}
\right.
\label{eq:MKPbilinear}
\end{eqnarray}
in which the subscripts of the bilinear operators correspond to
the components of the vector $\vec x=(x_1,x_2,\cdots,x_n)$,
while $\tau_0$ and $\tau_1$ are functions of $\vec x$.
By further putting some symmetry constraint on $\tau_0$ and $\tau_1$,
let us define as follows four (2+1)-dimensional PDEs
(see line ``2+1'' in Figure \ref{Fign+1}).

\begin{enumerate}

\item
With the system (\ref{eq:CKPbilinear}),
one defines by $D_4=0$ \cite{SH1982} the (2+1)-dim PDE labeled ``KP-1''
in Figure \ref{Fign+1}.

\item
With the system (\ref{eq:KPbilinear}),
one defines by $D_2=0$               the (2+1)-dim PDE labeled ``KP-2'' 
in Figure \ref{Fign+1}.

\item
With the system (\ref{eq:MKPbilinear})
and the $B_\infty$ symmetry constraint
\cite[p.~968]{JM1983}
\begin{eqnarray}
{\hskip -10.0 truemm}
& &
\left\lbrace
\begin{array}{ll}
\displaystyle{
\tau_0(x)=f(x_{\rm odd}) + x_2 g(x_{\rm odd})
+\frac{1}{2} x_2^2 h_1(x_{\rm odd}) + x_4 h_2(x_{\rm odd}) + \cdots,
}
\\
\displaystyle{
\tau_1(x)=f(x_{\rm odd}) - x_2 g(x_{\rm odd})
+\frac{1}{2} x_2^2 h_1(x_{\rm odd}) - x_4 h_2(x_{\rm odd}) + \cdots,
}
\end{array}
\right.
\label{constraintBinfty}
\end{eqnarray}
one defines the (2+1)-dim BKP equation
\begin{eqnarray}
& &
9 z_{x_1,x_5} - 5 z_{2 x_3}
\label{eq:BKP}
\\
& &
 + \bigg(
 z_{5x_1}
 +15 z_{x_1} z_{3 x_1}
 + 15\left(z_{x_1}\right)^3
 -5 z_{2x_1,x_3}- 15z_{x_1} z_{x_3}
\bigg)_{x_1} = 0,
\nonumber
\end{eqnarray}
in which $ z= \partial_{x_1}\Log \tau_0(\vec x)\vert_{x_2=x_4=\cdots =0}$
and $z_{2 x_3} \equiv z_{x_3x_3} \ldots$.

\item
With
(\ref{eq:KPbilinear})
and the $C_\infty$ symmetry constraint
\cite[p.~968]{JM1983}
\begin{eqnarray}
& &
\tau_0(x)=f(x_{\rm odd})
+\frac{1}{2} x_2^2 g(x_{\rm odd})
+\frac{1}{2} x_4^2 h(x_{\rm odd})
+ \cdots,
\label{constraintCinfty}
\end{eqnarray}
one defines the (2+1)-dim CKP equation 
\begin{eqnarray}
{\hskip -14.0 truemm}
& & 
9 z_{x_1,x_5} - 5 z_{2 x_3}
\label{eq:CKP}
\\
{\hskip -14.0 truemm}
& &
+ \bigg(
 z_{5x_1}
+ 15 z_{x_1} z_{3 x_1}
+ 15 \left(z_{x_1}\right)^3
-5 z_{2x_1,x_3}- 15 z_{x_1} z_{x_3} 
+\frac{45}{4}\left(z_{2x_1}\right)^2\bigg)_{x_1} = 0,
\nonumber
\end{eqnarray}
in which
$ z=\partial_{x_1}\Log \tau_0(\vec x)\vert_{x_2=x_4=\cdots =0} $.

\end{enumerate}

Next, from these (2+1)-dimensional PDEs,
one performs the following natural reductions to (1+1)-dimensional PDEs
(see line ``1+1'' in Figure \ref{Fign+1}).

\begin{enumerate}

\item
In KP-1,
the $C_\infty$ symmetry constraint (\ref{constraintCinfty}) defines
\begin{eqnarray}
& &
\left\lbrace
\begin{array}{ll}
\displaystyle{
\left( D_1^4 - 4 D_1 D_3\right) (f\cdot f) + 6 f g = 0,
}
\\
\displaystyle{
\left(D_1^3 + 2 D_3\right)(f\cdot g) = 0,\
}
\end{array}
\right.
\label{eq:bilinearHSII}
\label{eq:bilinearbiSH}
\end{eqnarray}
which we call \biSH\ \cite{SH1982} for reasons explained in next section.

\item
In KP-1,
the constraint
\begin{eqnarray}
{\hskip -10,0 truemm}
& &
\tau_0(x)=f(x_{\rm odd}) + x_2 g(x_{\rm odd})
+\frac{1}{2} x_2^2 h_1(x_{\rm odd}) + x_4 h_2(x_{\rm odd}) + \cdots,
\end{eqnarray}
defines
\begin{eqnarray}
& &
\left\lbrace
\begin{array}{ll}
\displaystyle{
\left(D_1^4 - 4 D_1 D_3\right) (f\cdot f) - 6 g^2 = 0,
}
\\
\displaystyle{
\left(D_1^3 + 2 D_3\right)(f\cdot g) = 0,\
}
\end{array}
\right.
\label{eq:c-KdV2}
\end{eqnarray}
which is called coupled KdV system of Hirota-Satsuma (HSS) \cite{SH1982}.

\item
In KP-2,
the elimination of $x_3$ \cite[p.~962]{JM1983}
yields the potential KdV$_5$ equation
\begin{equation}
\label{eq:potKdV5}
z_t + z_{xxxxx} + 5 z_{xx}^2 + 10 z_x z_{xxx} +10 z_x^3 = 0,
\end{equation}
with the notation $x \equiv x_1, t\equiv -x_5/16,z=2 \partial_x\Log\tau_0$.

\item
In BKP (\ref{eq:BKP}),
the reduction $z_{x_3}=0$ defines the potential SK equation \cite{SK1974}
\begin{eqnarray}
\label{eq:pSK}
z_t  + z_{xxxxx} + 15 z_x z_{xxx} + 15 \left(z_x\right)^3 = 0,
\end{eqnarray}
with the notation $x_5\equiv 9 t, x_1\equiv x$.

\item
In BKP (\ref{eq:BKP}),
the reduction $z_{x_5}=0$ defines
the 1+1-dimensional \biSK\ or Ramani equation \cite{Ramani1981}
\begin{eqnarray}
{\hskip -9.0 truemm}
\label{eq:Ramani}
&&\left(z_{xxxxx} + 15 z_x z_{xxx} + 15 \left(z_x\right)^3 - 15 z_x z_t
- 5 z_{xxt}\right)_x - 5 z_{tt} = 0,
\end{eqnarray}
with the notation $x_3\equiv t, x_1\equiv x$.

\item
In CKP (\ref{eq:CKP}),
the reduction $\partial_{x_3}\tau_0 = 0$ defines
the fifth order potential KK equation \cite{Kaup1980}
\begin{eqnarray}
\label{eq:pKK}
z_t  + z_{xxxxx} + 15 z_x z_{xxx} + 15 \left(z_x\right)^3
 + \frac{45}{4} \left(z_{xx}\right)^2 = 0,
\end{eqnarray}
with the notation $x_1\equiv x, x_5\equiv 9t$.

\item
In CKP (\ref{eq:CKP}),
the reduction $\partial_{x_5}\tau_0 = 0$ defines
the sixth order \biKK\ equation \cite{DyeParker2001}
\begin{eqnarray}
{\hskip -9.0 truemm}
& &
\left( z_{xxxxx} + 15 z_x z_{xxx} + 15 \left( z_x\right)^3 - 15
z_x z_t - 5 z_{xxt} + \frac{45}{4}\left(z_{xx}\right)^2\right)_x
\label{eq:pbiKK}
\\
{\hskip -9.0 truemm}
& &
- 5 z_{tt} = 0,
\nonumber
\end{eqnarray}
with the notation $x_1\equiv x , x_3\equiv t$.

\end{enumerate}

Finally,
the stationary reduction $(x,t) \to x-ct$ of these (1+1)-dimen\-sional PDEs
leads directly to the Hamiltonian systems or the ODE
listed in the line ``0+1'' of Figure \ref{Fign+1}.

The four ODEs F-III, F-IV, F-V, F-VI have a singlevalued general solution,
obtained by the Jacobi postmultiplier method \cite{Cos2000a},
which is expressed with genus-two hyperelliptic functions.
Three of them (F-III, F-IV, F-V),
which are the stationary reductions of
respectively
(\ref{eq:pKK}),
(\ref{eq:pSK}),
and
(\ref{eq:potKdV5}),
have been shown \cite{Fordy1991}
to have a one-to-one correspondence with the $q_1$ component of
the three integrable cases of HH3.
Therefore
the chain of reductions generated from the systems
(\ref{eq:KPbilinear})
and
(\ref{eq:MKPbilinear})
contains the full information for the integration of HH3.

Let us now show that Figure \ref{Fign+1}
also contains the full information for the integration of HH4.
This will involve two kinds of coupled KdV (\cKdV) systems:
some with a fourth order Lax pair,
some with a fifth order Lax pair.

\section{Link of coupled KdV systems with HH4}

In the variables
$u=\partial_x^2 \Log f,\ v= 4 g/f$,
the bilinear system (\ref{eq:bilinearbiSH}) is rewritten as
the \cKdV\ system \cite{DS1981,SH1982,JM1983,DS1984}
\begin{eqnarray}
& &
\left\lbrace
\begin{array}{ll}
\displaystyle{
-4 u_t + \left(6 u^2 + u_{xx} + 3 v\right)_x = 0,
}
\\
\displaystyle{
2 v_t + 6 u v_x + v_{xxx} = 0,\ \ \
}
\end{array}
\right.
\label{eq:systemHSII}
\end{eqnarray}
with the notation $x_3\equiv t, x_1\equiv x$.
This system possesses the fourth order Lax pair
\cite{SH1982}
\begin{eqnarray}
& &
\left\lbrace
\begin{array}{ll}
\displaystyle{
\left(\partial_x^4+4u\partial_x^2+4u_x\partial_x+2u_{xx}+
4u^2+v\right)\psi\ =\lambda\psi,
}
\\
\displaystyle{
\left(\partial_x^3+3u\partial_x+\frac{3}{2}u_x\right)\psi\
=\partial_t\psi,
}
\end{array}
\right.
\label{eq:xLaxHS2}
\end{eqnarray}

\begin{figure}[h]
\centering\epsfig{file=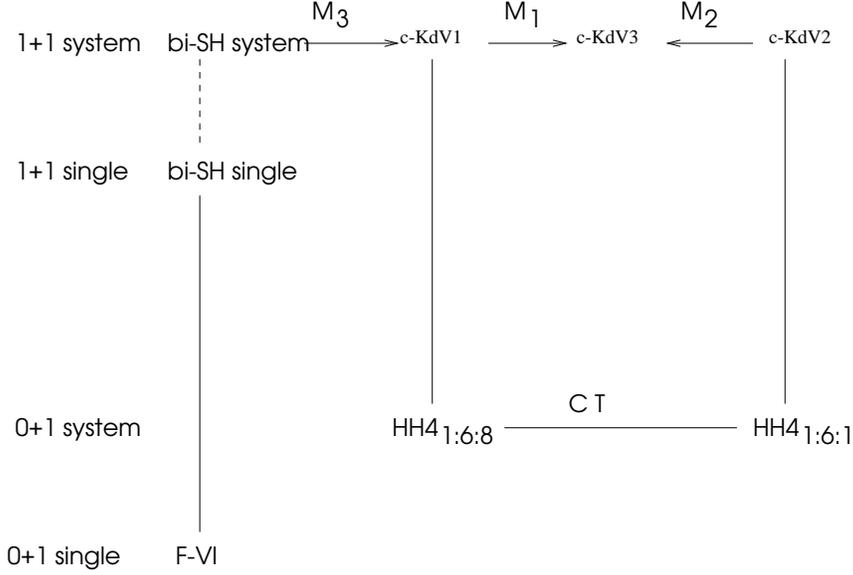,width=0.9\textwidth}
\caption{
Path from an already integrated ODE (autonomous F-VI)
to the quartic cases 1:6:1 and 1:6:8.
All 1+1-dimensional systems involved (on the top line)
have fourth order Lax pairs.
The dashed vertical line from the level ``1+1-system'' 
to the level ``1+1-single''
represents the elimination of one dependent variable.
All the other vertical lines represent the stationary reduction.
The horizontal lines represent
Miura transformations at the level ``1+1-system''
and 
canonical transformations at the Hamiltonian level ``0+1-system''.
The systems are defined as 
(\ref{eq:cKdV1})       for \cKdV$_1$,
(\ref{eq:c-KdV2})      for \cKdV$_2$, 
[1, p.~79]             for \cKdV$_3$,                          
(\ref{eq:systemHSII})  for the \biSH\ system.
The Miura maps M$_1$, M$_2$ can be found in 
[1, Eq.~(5.3)]             and                                 
[1, Eq.~(5.8)].                                                
}\label{FigLax4}
\end{figure}

Under the Miura transformation denoted $M_3$ in Figure \ref{FigLax4}
\begin{eqnarray}
{\hskip -10.0 truemm}
& &
\left\lbrace
\begin{array}{ll}
\displaystyle{
4 u=2G-F_x-F^2,
}
\\
\displaystyle{
2 v= 2F_{xxx}+4FF_{xx}+8GF_x+4FG_x+3F_x^2-2F^2F_x-F^4+4GF^2,
}
\end{array}
\right.
\end{eqnarray}
the system (\ref{eq:systemHSII})
is mapped to the following \cKdV\ system
(denoted \cKdV${}_1$ in Figure \ref{FigLax4})
given in \cite{BEF1995b,BakerThesis}
\begin{eqnarray}
{\hskip -10.0 truemm}
& &
\left\lbrace
\begin{array}{ll}
\displaystyle{
4 F_t=\left(-2F_{xx}-3FF_{x}+F^3-6FG\right)_x,
}
\\
\displaystyle{
8 G_t= 2G_{xxx}+12GG_x+6 FG_{xx}+12GF_{xx}+18F_xG_x-6F^2G_x
}
\\
\displaystyle{
\phantom{xxxxx}+3F_{xxxx} +3FF_{xxx}+18F_xF_{xx}-6F^2F_{xx}-6FF_x^2,
}
\end{array}
\right.
\label{eq:cKdV1}
\end{eqnarray}
with the Lax pair
\begin{eqnarray}
& &
\left\lbrace
\begin{array}{ll}
\displaystyle{
\left(\partial_x^4 + (2 G - F_x - F^2) \partial_x^2
 + (2 G - F_{x} - F^2)_x \partial_x
\right.
}
\\
\displaystyle{
\left.+(F G)_x + G_{xx} + G^2\right) \psi=\lambda \psi,
}
\\
\displaystyle{
\left(\partial_x^3 + \frac{3}{4} (2 G - F_x - F^2) \partial_x
 + \frac{3}{8} (2 G - F_{x} - F^2)_x \right) \psi = \partial_t \psi,
}
\end{array}
\right.
\end{eqnarray}

The stationary reduction of this \cKdV${}_1$ system happens to be
the case 1:6:8 of HH4 for arbitrary values of $(\beta,\mu)$.

The field $z=\int u \D x$ of (\ref{eq:systemHSII})
satisfies the sixth order PDE \cite{Bog1990},
\begin{eqnarray}
-8 z_{tt} + z_{xxxxxx} - 2 z_{xxxt} + 18 z_x z_{xxxx} + 36 z_{xx}z_{xxx}
+ 72 z_x^2 z_{xx} = 0\ ,
\label{eq:HSII}
\end{eqnarray}
which is of second order in time and
which for this reason we call bidirectional Satsuma-Hirota (\biSH) equation.
Its stationary reduction is identical to the autonomous case of the F-VI
nonlinear ODE,
integrated \cite{Cos2000a} with genus-two hyperelliptic functions.

Therefore, since there exists a path from the
 (not yet integrated in its full generality)
1:6:8 HH4 case and the (integrated) autonomous F-VI ODE,
the general solution of the 1:6:8 can in principle be obtained,
this will be addressed in future work.

All the links between the system
(\ref{eq:systemHSII}) and other \cKdV\ systems
considered by S.~Baker and which reduce to the integrable cases
1:6:1 and 1:6:8 of HH4 are displayed in Figure \ref{FigLax4}.

\begin{figure}[h]
\centering\epsfig{file=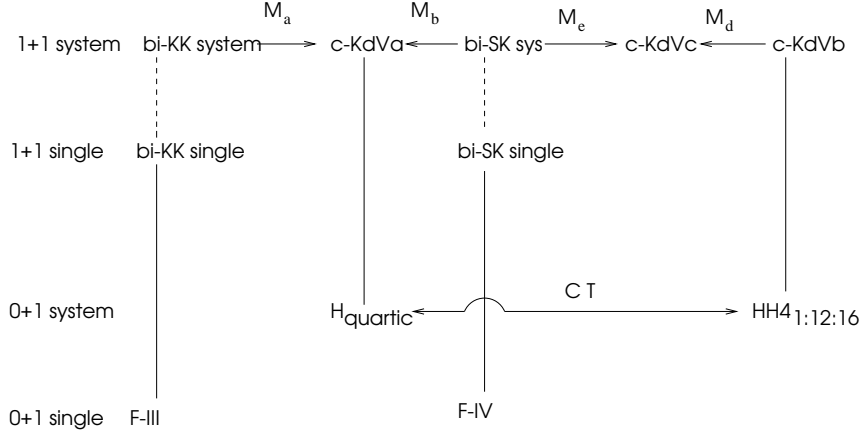,width=0.9\textwidth}
\caption{
Path from an already integrated ODE (autonomous F-III or F-IV,
which are ODEs for $q_1$ in the HH3-KK and HH3-SK cases)
to the quartic case 1:12:16.
All 1+1-dimensional systems involved (on the top line)
have fifth order Lax pairs.
The dashed vertical line from the level ``1+1-system'' 
to the level ``1+1-single''
represents the elimination of one dependent variable.
All the other vertical lines represent the stationary reduction.
The horizontal lines represent
Miura transformations at the level ``1+1-system''
and birational canonical transformations at the Hamiltonian level ``0+1-system''.
The Miura maps $M_a,M_b$ are given in the text,
M$_d$ is given in [1, p.~95],                                 
$M_e=M_b M_c$,
in which $M_c$ is the Miura transformation from \cKdV a to \cKdV c
given in [1, p.~95].                                          
The systems are defined as 
[1, Eq.~(6.9)] for \cKdV b,                                   
and
[1, p.~95]     for \cKdV c.                                   
}
\label{FigLax5}
\end{figure}

Finally,
let us explain the link between the 1:12:16 integrable case of HH4
and two \cKdV\ systems possessing a fifth order Lax pair,
systems respectively equivalent to
the \biSK\ equation  (\ref{eq:Ramani}) and
the \biKK\ equation (\ref{eq:pbiKK}).

The following coupled system \cite{DS1984}
\begin{eqnarray}
& &
\left\lbrace
\begin{array}{ll}
\displaystyle{
u_t = \left(-2 a u_{xx} - b u^2 +\frac{9 a^2}{5 b} v\right)_x,
}
\\
\displaystyle{
v_t = a v_{xxx}- b u_{xxxxx} -\frac{5 b^2}{3 a} u u_{xxx}
- \frac{5 b^2}{3 a} u_x u_{xx} + b u v_x - b u_x v,
}
\end{array}
\right.
\label{eq:cRamani}
\end{eqnarray}
where $a,b$ are nonzero constants, arises from the
compatibility condition of the fifth order Lax pair
\begin{eqnarray}
& &
\left\lbrace
\begin{array}{ll}
\displaystyle{
\left(\partial_x^5 + \frac{5 b}{3 a} u \partial_x^3
 + \frac{5 b}{3 a} u_x \partial_x^2 + v \partial_x\right) \varphi
 = \lambda\varphi,
}
\\
\displaystyle{
\left(a\partial_x^3 + b u \partial_x\right)\varphi= \partial_t\varphi.
}
\end{array}
\right.
\label{eq:Lax-Ramani}
\end{eqnarray}
The field $z=\int u \D x$ of (\ref{eq:cRamani})
satisfies the sixth order PDE
\begin{eqnarray}
\label{eq:Ramani-ab}
&& 5 z_{tt} +\left( 5 z_{xxt} + 5 \frac{b}{a} z_t z_x - a
 z_{xxxxx} - 5 b z_x z_{xxx} - \frac{5 b^2}{3a} z_x^3\right)_x=0,
\end{eqnarray}
identical to the \biSK\ equation (\ref{eq:Ramani}) for $a=1,\ b=3$.

Similarly,
the coupled system \cite{DS1984}
\begin{eqnarray}
& &
\left\lbrace
\begin{array}{ll}
\displaystyle{
u_t = \left(-\frac{7}{2} a u_{xx} - b u^2 +\frac{9 a^2}{5 b} v\right)_x,
}
\\
\displaystyle{
v_t =\frac{5}{2} a v_{xxx}-\frac{19}{4} b u_{xxxxx}
-\frac{25 b^2}{6a}  u u_{xxx}
- \frac{5 b^2}{a} u_x u_{xx} + b u v_x - b u_x v,
}
\end{array}
\right.
\label{eq:cbiKK}
\end{eqnarray}
arises from the compatibility condition of the other fifth order Lax pair
\begin{eqnarray}
& &
\left\lbrace
\begin{array}{ll}
\displaystyle{
\left(\partial_x^5 + \frac{5 b}{3 a} u \partial_x^3
+ \frac{5 b}{2 a} u_x \partial_x^2 + v \partial_x + \frac{1}{2}v_x
- \frac{5b}{12a} u_{xxx}\right)\varphi =\lambda\varphi,
}
\\
\displaystyle{
\left(a\partial_x^3 + b u \partial_x +\frac{b}{2}u_x\right)\varphi=
\partial_t\varphi.
}
\end{array}
\right.
\label{eq:Lax-biKK}
\end{eqnarray}

The field $z=\int u \D x$ satisfies the sixth order PDE
\begin{equation}
\label{eq:pbiKK-ab}
5 z_{tt} + a\left( 5 z_{xxt} + 5 \frac{b}{a} z_t z_x - a
 z_{xxxxx} - 5 b z_x z_{xxx} - \frac{5 b^2}{3a} z_x^3
-\frac{15 b}{4}  z_{xx}^2\right)_x=0,
\end{equation}
identical to the potential \biKK\ equation (\ref{eq:pbiKK}) for $a=1, b=3$.
The property of these two systems which is of interest to us is
the existence of two mappings,
respectively (setting $a=5$),
for the system (\ref{eq:cRamani}) the Miura transformation
denoted $M_{\rm b}$                                          
\begin{eqnarray}
& &
\left\lbrace
\begin{array}{ll}
\displaystyle{
u_{\rm \biSK}=\frac{3}{b}(2 G + 3 F_x - F^2)\ ,
}
\\
\displaystyle{
v_{\rm \biSK}=F_{xxx} + G_{xx} - F F_{xx} + G F_x - F G_x  + G^2\ ,
}
\end{array}
\right.
\end{eqnarray}
and, for the system (\ref{eq:cbiKK}),
the transformation denoted $M_{\rm a}$                       
\begin{eqnarray}
& &
\left\lbrace
\begin{array}{ll}
\displaystyle{
u_{\rm \biKK}=\frac{3}{b}(2 G - 2 F_x - F^2),
}
\\
\displaystyle{
v_{\rm \biKK}=-F_{xxx} + 3 G_{xx} - F F_{xx} + 2 F G_x - F_x^2 + G^2,
}
\end{array}
\right.
\end{eqnarray}
to a common coupled KdV-type system \cite[p.~65]{BakerThesis}
(denoted \cKdV$_a$ in Figure \ref{FigLax5})

\begin{eqnarray}
{\hskip -10.0 truemm}
& &
\left\lbrace
\begin{array}{ll}
\displaystyle{
F_t= \left(- 7 F_{xx} - 3 G_{x} - 3 F F_{x} - 9 F G + 2 F^3\right)_x,
}
\\
\displaystyle{
G_t= 3 F_{xxxx} + 2 G_{xxx} + 3 F G_{xx}- 3 F^2 F_{xx} - 3 F^2 G_x -
3 F F_x^2 
}
\\
\displaystyle{
\phantom{G_t=}
+ 3 F G F_x + 9 F_x F_{xx} + 9 F_x G_x + 3 G F_{xx} + 3 G G_x.
}
\end{array}
\right.
\label{eq:cKdVunphysical}
\end{eqnarray}

This system also possesses a fifth order Lax pair,
which can be written in two different ways, either
\begin{eqnarray}
& &
\left\lbrace
\begin{array}{ll}
\displaystyle{
(\partial_x^2 +F\partial_x + F_x + G)\partial_x (\partial_x^2 - F
\partial_x +G)\varphi = \lambda\varphi,
}
\\
\displaystyle{
\left(5 \partial_x^3 + 3(2 G -2F_x - F^2)\partial_x + 3(G_x-F_{xx}- F F_x)
\right) \varphi =\partial_t\varphi
}
\end{array}
\right.
\label{Lax1:SB}
\end{eqnarray}
or
\begin{eqnarray}
& &
\left\lbrace
\begin{array}{ll}
\displaystyle{
(\partial_x^2 - F
\partial_x +G) (\partial_x^2 +F\partial_x + F_x + G)\partial_x \varphi
 = \lambda\varphi,
}
\\
\displaystyle{
\left( 5 \partial_x^3 + 3(2 G +3 F_x - F^2)\partial_x\right)\varphi
= \partial_t\varphi.
}
\end{array}
\right.
\label{Lax2:SB}
\end{eqnarray}

It happens that the stationary reduction of (\ref{eq:cKdVunphysical}),
which is an unphysical Hamiltonian system \cite[pp.~98, 103]{BakerThesis},
is mapped by a canonical transformation to the 1:12:16 case of HH4.

In Figure \ref{FigLax5},
we display the link between \cKdV\ systems possessing a fifth order Lax pair
and the 1:12:16 integrable Hamiltonian.

\section{Conclusion}

We have linked each of the three not yet integrated quartic H\'enon-Heiles 
cases to fourth order ODEs recently integrated by Cosgrove,
\textit{via} a path involving,
on one hand
canonical transformations between Hamiltonian systems,
and on the other hand
B\"acklund transformations between coupled KdV systems.
This proves that these three cases have a general solution expressed with
hyperelliptic functions of genus two.
Their explicit closed form expression will be given in future work.

\section{Acknowledgements}

The authors acknowledge the financial support of the Tournesol grants 
T99/040 and T2003.09.
MM and RC thank the organizers of this ARW for invitation.
CV is a research assistant of the FWO.


\begin{thebibliography}{99}

\bibitem{BakerThesis} S.~Baker,
Squared eigenfunction representations of integrable hierarchies,
PhD Thesis, University of Leeds (1995).

\bibitem{BEF1995b} S.~Baker, V.~Z.~Enol'skii, and A.~P.~Fordy,
Integrable quartic potentials and coupled KdV equations,
Phys.~Lett.~A {\bf 201} (1995) 167--174.

\bibitem{Bog1990} O.~I.~Bogoyavlensky,
Breaking solitons in $2+1-$dimensional integrable equations,
Usp.~Matem.~Nauk {\bf 45} (1990) 17--77
[English~: Russ.~Math.~Surveys {\bf 45} (1990) 1--86].

\bibitem{BSV:1982} T.~Bountis, H.~Segur and F.~Vivaldi,
Integrable Hamiltonian systems and the Painlev\'e property,
Phys.~Rev.~A {\bf 25}, 1257--1264 (1982).

\bibitem{CTW} Chang Y.~F., M.~Tabor, and J.~Weiss,
Analytic structure of the H\'enon-Heiles Ha\-miltonian in integrable and
nonintegrable regimes,
J.~Math.~Phys.~{\bf 23} (1982) 531--538.

\bibitem{Cos2000a} C.~M.~Cosgrove,
Higher order Painlev\'e equations in the polynomial class,
I. Bureau symbol $P2$,
Stud.~Appl.~Math.~{\bf 104} (2000) 1--65.

\bibitem{Drach1919KdV} J.~Drach,
Sur l'int\'egration par quadratures de l'\'equation
${\D^2 y \over \D x^2} = [\varphi(x) + h] y,$
\CRAS\ {\bf 168} (1919) 337--340.

\bibitem{DS1981} V.~G.~Drinfel'd and V.~V.~Sokolov,
Equations of Korteweg-de Vries type and simple Lie Algebras,
Soviet Math.~Dokl.~{\bf 23} (1981) 457--462.

\bibitem{DS1984} V.~G.~Drinfel'd and V.~V.~Sokolov,
Lie algebras and equations of Korteweg-de Vries type,
Itogi Nauki i Tekhniki, Seriya Sovremennye Problemy Matematiki
{\bf 24} (1984) 81--180
[English: Journal of Soviet Math.~{\bf 30} (1985) 1975--2036].

\bibitem{DyeParker2001} J.~M.~Dye and A.~Parker,
On bidirectional fifth-order nonlinear evolution equations, Lax
pairs, and directionally dependent solitary waves,
J.~Math.~Phys.~{\bf 42} (2001) 2567--2589.

\bibitem{Fordy1991} A.~P.~Fordy,
The H\'enon-Heiles system revisited,
Physica D {\bf 52} (1991) 204--210.

\bibitem{GDP1982b} B.~Grammaticos, B.~Dorizzi, and R.~Padjen,
Painlev\'e property and integrals of motion for the H\'enon-Heiles system,
Phys.~Lett.~A {\bf 89} (1982) 111--113.

\bibitem{GDR1983} B.~Grammaticos, B.~Dorizzi, and A.~Ramani,
Integrability of Hamiltonians with third- and fourth-degree polynomial
potentials,
J.~Math.~Phys.~{\bf 24} (1983) 2289--2295.

\bibitem{HH} M.~H\'enon and C.~Heiles,
The applicability of the third integral of motion: some numerical
experiments,
Astron.~J.~{\bf 69} (1964) 73--79.

\bibitem{H1987} J.~Hietarinta,
Direct method for the search of the second invariant,
Phys.~Rep.~{\bf 147} (1987) 87--154.

\bibitem{JM1983} M.~Jimbo and T.~Miwa,
Solitons and infinite dimensional Lie algebras,
Publ.~RIMS, Kyoto {\bf 19} (1983) 943--1001.

\bibitem{Kaup1980} D.~J.~Kaup,
On the inverse scattering problem for cubic eigenvalue problems of the class
$ \psi_{xxx} + 6 Q \psi_ x + 6 R \psi  = \lambda  \psi $,
Stud.~Appl.~Math.~{\bf 62} (1980) 189--216.

\bibitem{Manakov1973} S.~V.~Manakov,
On the theory of two-dimensional stationary self-focusing of electromagnetic
waves,
{\em Zh.~Eksp.~Teor.~Fiz.~} {\bf 65} (1973) 505--516
[JETP {\bf 38} (1974) 248--253].

\bibitem{Ramani1981} A.~Ramani,
in {\it Fourth international conference on collective phenomena}, 
ed.~J.~L.~Lebowitz,
Annals of the New York Academy of Sciences {\bf 373} 54--67
(NY Ac.~Sc., New York, 1981).

\bibitem{RDG1982} A.~Ramani, B.~Dorizzi, and B.~Grammaticos,
Painlev\'e conjecture revisited,
Phys.~Rev.~Lett.~{\bf 49} (1982) 1539--1541.
  
\bibitem{SH1982} J.~Satsuma and R.~Hirota,
A coupled KdV equation is one case of the four-reduction of the KP hierarchy,
J.~Phys.~Soc.~Japan~{\bf 51} (1982) 3390--3397.

\bibitem{SK1974} K.~Sawada and T.~Kotera,
A method for finding N-soliton solutions of the K.d.V.~equation and
K.~d.~V.-like equation,
\PTP\ {\bf 51} (1974) 1355--1367.

\bibitem{V2003} C.~Verhoeven,
Integration of Hamiltonian systems of H\'enon-Heiles type
and their associated soliton equations,
PhD thesis, Vrije Universiteit Brussel (28 May 2003).

\bibitem{VM2003} C.~Verhoeven and M.~Musette,
Soliton solutions of two bidirectional sixth order partial differential
equations belonging to the KP hierarchy,
J.~Phys.~A {\bf 36} (2003) L133--L143.

\bibitem{VMC2002} C.~Verhoeven, M.~Musette and R.~Conte,
\hfill \break \noindent                                       
Integration of a generalized H\'enon-Heiles Hamiltonian,
J.~Math.~Phys.~{\bf 43} (2002) 1906--1915.
http://arXiv.org/abs/nlin.SI/0112030.

\bibitem{VMC2003} C.~Verhoeven, M.~Musette and R.~Conte,
General solution for Hamiltonians with extended cubic and quartic potentials,
                        Theor.~Math.~Phys.~{\bf 134} (2003) 128--138.
http://arXiv.org/abs/nlin.SI/0301011.

\bibitem{Woj1985} S.~Wojciechowski,
Integrability of one particle in a perturbed central quartic potential,
Physica Scripta {\bf 31} (1985) 433--438.

\end{thebibliography}
\end{document}